\documentclass[conf]{new-aiaa}
\usepackage[utf8]{inputenc}

\usepackage{graphicx}
\usepackage{soul}
\usepackage{amsmath}
\usepackage{bm}
\usepackage[version=4]{mhchem}
\usepackage{siunitx}
\usepackage{longtable,tabularx}

\title{Line-of-Sight Deep-Space Autonomous Navigation}

\author{V. Franzese\footnote{PhD Candidate, Dept. of Aerospace Science and Technology, Politecnico di Milano, Via La Masa, 34, Milan, Italy, vittorio.franzese@polimi.it.}, F. Topputo \footnote{Associate Professor, Dept. of Aerospace Science and Technology, Politecnico di Milano. Member AIAA. francesco.topputo@polimi.it}}
\affil{Politecnico di Milano, Via La Masa 34, 20156, Milan, Italy}

\begin{document}

\maketitle

\begin{abstract}
Autonomous navigation is one of the main enabling technologies for future space missions. While conventional spacecraft are navigated through ground stations, their employment for deep-space CubeSats yields costs comparable to those of the platform.  This paper introduces an extended Kalman filter formulation for spacecraft navigation exploiting the line-of-sight observations of visible Solar System objects to infer the spacecraft state. The line-of-sight error budget builds up on typical performances of deep-space CubeSats and includes uncertainties deriving from the platform attitude, the image processing, and the sensors performances. The errors due to the low-thrust propagation and light-time delays to the navigation beacons are also taken into account. Preliminary results show feasibility of the deep-space autonomous navigation exploiting the line-of-sight directions to visible beacons with a 3$\sigma$ accuracy of 1000 km for the position components and 2 m/s for the velocity components.
\end{abstract}

\section{Introduction}
There is a recent growing interest in miniaturized interplanetary spacecraft \cite{poghosyan2017cubesat}: ESA has financed several interplanetary CubeSat mission studies (M-ARGO \cite{walker2018deep}, LUMIO \cite{topputo2018lumio, Speretta2019, cipriano2018orbit}, VMMO \cite{VMMO2018cospar}, CubeSats along the AIM/HERA mission \cite{michel2018hera}); NASA funded 19 SmallSat deep-space mission studies after MarCO \cite{klesh2015marco}, the first interplanetary CubeSat launched along with Insight mission.

Nanosatellites, or CubeSats, may lower the absolute entry-level cost by an order of magnitude, still returning useful science/exploration objectives. Moreover, they can be built and assembled using COTS components, and can share the launch with either a main passenger or other CubeSats. While the costs to design and build a CubeSat scale with size, those needed to operate it do not follow this trend in case ground-based navigation is used. This is because ground station and navigation teams would still be needed like in a conventional spacecraft. Most of the deep-space navigation techniques rely on radiometric tracking and orbit determination through ground-based tracking stations \cite{Thornton2003}. This practice is the state of the art for Earth-orbiting satellites and deep-space probes. Radiometric measurements yield position accuracy in the order of meters in low Earth orbit and kilometres in deep-space. The drawback relies in the interaction with the ground station, which is unavoidable. For every mission and mission phase, a dedicated Flight Dynamics team has to be allocated, so contributing significantly to the overall ground segment costs. Navigating a large satellite or a CubeSat is the same from the ground-stations point of view. Moreover, the signal delays between ground stations and deep-space probes constraints spacecraft operations as well. Automation is required for next-generation missions \cite{quadrelli2015guidance}. 

Partial or full autonomous navigation methods are becoming more and more appealing within the space community and in particular for interplanetary CubeSats due to both the considerable savings in the overall missions cost and the possibility of managing multiple missions at once. The output of a navigation method is the determination of a spacecraft position, state, or orbit via exploitation of some kind of measurements \cite{schutz2004statistical}. Among the autonomous navigation methods, the X-ray pulsar navigation records a pulsar signal and computes the time-of-arrival difference with respect to the  Solar System barycenter to estimate the relative spacecraft range \cite{Sheikh2006, Anderson2015}, the horizon-based navigation exploits the apparent full-disk of known spherical or nearly ellipsoidal bodies and the attitude knowledge to estimate the spacecraft position vector relative to the body \cite{franzese2019autonomous, Mortari2016, holt2018, Christian2016_December, Christian2016_May, Christian2017}, and the navigation using visible objects \cite{Mortari2017, Karimi2015} employs the line-of-sight (LOS) directions to some bodies for which ephemerides are known to determine the spacecraft position. When adopting optical measurements of distant objects which are seen as light dots, the navigation solution can be computed and then corrected for the light-time delay effect.

In this work, a position determination method and a consequent Kalman filter implementation which exploit the line-of-sight directions to navigation beacons are presented. The light-time delay phenomenon is directly taken into account in the navigation method rather than considered a posteriori. The LOS directions are affected by the typical uncertainties of CubeSat sensors.
A peculiar measurement strategy of one-object-at-time tracking is then defined to comply with the limited number of camera on board a CubeSat.  The position and orbit determination methods are described in Section \ref{sec:Methodology}, while the performances of the methods are reported in Section \ref{sec:Performances}.


\section{Methodology} \label{sec:Methodology}
\subsection{Geometry of the problem}
\begin{figure}[ht]
	\centering
	\includegraphics[width=0.4\columnwidth]{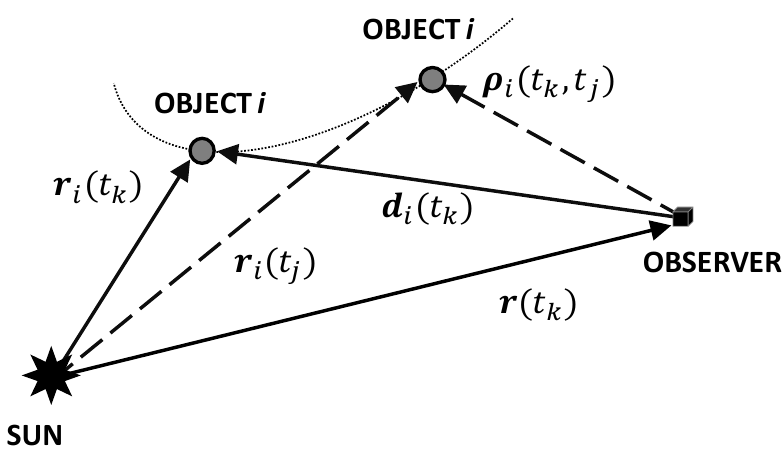}
	\caption{Geometry of the observation problem.}
	\label{fig:Geometry}
\end{figure}

The geometry of a spacecraft which is observing a distant object is shown in Figure \ref{fig:Geometry}. The spacecraft position vector in an inertial frame and at a given epoch $t_k$ is denoted as $\bm{r}(t_k)$, the object position vector at the same epoch is represented as $\bm{r}_{i}(t_k)$, and the geometric position vector of the object with respect to the spacecraft as $\bm{d}_i(t_k)$.
What is truly observed by the spacecraft, however, is not the geometric position of the object but its apparent one. This is due to the time required for the light to depart from the object and arrive to the spacecraft location. This phenomenon is known as light-time delay. Thus, the light is coming from the object position at a previous epoch, denoted as $\bm{r}_{i}(t_j)$. The object apparent position vector relative to the observer is then $\bm{\rho}_i(t_k, t_j)$.
 \\
With reference to Figure \ref{fig:Geometry}, the spacecraft position can be written as 
\begin{equation} \label{eq:r_tk}
\bm{r}(t_k) = \bm{r}_{i}(t_j) - \bm{\rho}_i(t_k, t_j).
\end{equation}
The apparent object position vector can be written as its modulus and line-of-sight direction. The modulus depends on the time required for the light to depart from the object at the epoch $t_j$ and arrive to the spacecraft at $t_k$, while the line-of-sight is measured by a camera or a star tracker. Thus, considering the light time delay as $\delta t_i = t_k - t_j$, the object apparent position vector can be rewritten as
\begin{equation} \label{eq:lighttime}
    \bm{\rho}_i = c \, \delta t_i \, \hat{\bm{\rho}}_i 
\end{equation}
where c is the speed of light and $\hat{\bm{\rho}}_i$ the object apparent line-of-sight. 
\subsection{Position Determination} \label{subsec:PosDet}
The spacecraft is now assumed to observe $n$ objects. In order to establish a minimization procedure and observing that $t_j = t_k - \delta t_i$, Equation \ref{eq:r_tk} can be written in the residual form as
\begin{equation} \label{eq:residualform}
    \bm{\varepsilon}_i = \bm{r}(t_k) - \left[ \bm{r}_i(t_k - \delta t_i) - c \, \delta t_i \, \bm{\hat{\rho}}_{i} \right],
\end{equation}
where $\bm{\epsilon}_i$ is the residual vector. This is used to define the cost function
\begin{equation} \label{eq:costfcn_posdet}
    J = \sum_{i = 1}^n \bm{\varepsilon}_i^T \, \bm{\varepsilon}_i.
\end{equation}
The inputs of the position determination process are the epoch of observation $t_k$, the line-of-sight directions $\bm{\hat{\rho}}_i$ to the $n$ bodies at the epoch $t_k$, and the ephemeris of the bodies. The variables to determine are then $\bm{r}(t_k)$ and the light time delays $\delta t_i$ to the objects, for a total of $3+i$ variables. Minimization procedures can exploit the cost function gradient, which in this case is a $(3+i)$ length vector whose components are
\begin{equation} 
    \begin{aligned}
    \frac{\partial{J}}{\partial{\bm{r}(t_k)}} = & \, 2 \sum_{i=1}^n \, \bm{\varepsilon}_{i} \, , & \\
    \frac{\partial{J}}{\partial{\delta t_i}} = & \, 2 \, \left[ \bm{v}_i(t_k - \delta t_i) + c \, \bm{\hat{\rho}}_i\right]^T  \bm{\varepsilon}_{i}, & \forall \, i = 1,..,n,\\
    \end{aligned}
\end{equation}
where $\bm{v}_i(t_k - \delta t_i)$ is the velocity of the $i$-esimal object at the epoch $t_j$, known by ephemeris models. The minimization of Equation \ref{eq:r_tk} can be performed iteratively with a Newton-like method. This method solves for the instantaneous position of the spacecraft at a given epoch and the light-time delays to the observed bodies.
\subsection{Kalman filter implementation}
The position determination method described in Section \ref{subsec:PosDet} solves for the spacecraft position and the related light-time delays to the observed bodies at a defined epoch. Thus, in order link the solution to a given orbital dynamics and to estimate also the spacecraft velocity, the LOS measurements can be included into an extended Kalman filter.  
\subsubsection{The process}
In this formulation, the spacecraft state can be defined as 
\begin{equation}
    \bm{x} = [\bm{r} \, \, \bm{v} \, \, \bm{\delta t}] \, ^T
\end{equation}
where $\bm{r}$ is the spacecraft position vector, $\bm{v}$ the spacecraft velocity vector, and $\bm{\delta t}$ is a vector whose components are the light-time delays $\delta t_i$ to the observed bodies. The state vector dimension is ($6+i$). The state derivative is then 
\begin{equation} \label{eq:dynamics}
    \bm{\dot{x}} = \bm{f}(\bm{x}, t) = [\bm{v} \, \, \bm{a} \, \, \bm{\dot{\delta t}}] \, ^T
\end{equation}
where $\bm{a}$ is the spacecraft acceleration and $\bm{\dot {\delta t}}$ the rate of change of the light time delays with respect to the observed bodies. The spacecraft acceleration is given by the Sun gravity only during coasting arcs ($\bm{a}_C$), while it is also affected by the low-thrust during the thrust arcs ($\bm{a}_T$), so that
\begin{equation}
    \bm{a}_C = - \mu_S \frac{\bm{r}}{\lVert \bm{r} \rVert^3} \qquad \qquad \bm{a}_T = - \mu_S \frac{\bm{r}}{\lVert \bm{r} \rVert^3} + \bm{f}_T 
\end{equation}
where $\mu_S$ is the Sun gravity constant and $\bm{f}_T$ the low-thrust acceleration. The components of $\bm{\dot {\delta t}}$ derive from Equation \ref{eq:r_tk} and Equation \ref{eq:lighttime} as 
\begin{equation}
    \delta t_i = \frac{\lVert \bm{\rho}_i \rVert}{c} = \frac{\lVert \bm{r}_i(t-\delta t_i) - \bm{r}(t) \rVert}{c} \qquad \rightarrow \qquad \dot{\delta t_i} = \frac{(\bm{v}_i(t-\delta t_i)-\bm{v}(t))^T \bm{\hat{\rho}}_i}{c} 
\end{equation}
\subsubsection{The measurements}
The measurements given to the Kalman filter are the azimuth and elevation angles ($\theta$ and $\phi$) coming from the apparent line-of-sight directions $\bm{\hat {\rho}}_i$ to the $n$ objects. Thus, the measurement equation is 
\begin{equation} \label{eq:measurements}
    \bm{y} = \bm{h} (\bm{x}, t) = [\bm{y}_1, \bm{y}_2,..., \bm{y}_i,..., \bm{y}_n]^T;  
    \end{equation}
    where
\begin{equation}
    \bm{y}_i =  \begin{bmatrix} \theta_i \\
                                \phi_i 
                \end{bmatrix}
                =
                \begin{bmatrix}
                        \textrm{atan}\left(\frac{\bm{\hat{\rho}}_{i,y}}{\bm{\hat{\rho}}_{i,x}}\right) \\
                        \textrm{asin}\left(\bm{\hat{\rho}}_{i,z}\right)
                \end{bmatrix}
\end{equation}
where the first, the second, and the third components of the LOS directions are $\bm{\hat {\rho}}_{i,x}$, $\bm{ \hat {\rho}}_{i,y}$, and $\bm{ \hat {\rho}}_{i,z}$, respectively. 
\subsubsection{The algorithm}
The extended Kalman filter (EKF) equations in the continuous-discrete form \cite{crassidis2004optimal, Simon2006} are shown in Algorithm \ref{Alg:ExtendedKalmanFilter}, where $\bm{x}$ is the state vector, $\bm{f}$ comes from Equation \ref{eq:dynamics}, $\bm{h}$ comes from Equation \ref{eq:measurements}, $\bm{P}$ is the state error covariance, $\bm{F}$ is $\partial\bm{f}/\partial\bm{x}$, $\bm{H}$ is $\partial\bm{h}/\partial\bm{x}$, $\bm{K}$ is the Kalman gain, and $k$ represents a given time step. The process and measurement errors, $\bm{w}$ and $\bm{v}$, respectively, are mutually uncorrelated zero-mean white noise processes with known covariances ($\bm{Q}$ and $\bm{R}$, respectively). The "$\string^$" superscript denotes estimated quantities, while "$+$" and "$-$" are used for corrected and predicted quantities, respectively. The filter settings are presented in Section \ref{subsec:kalmanfiltersettings}.
\begin{algorithm}[ht]
\setstretch{1.25}
\caption{\textbf{\quad Extended Kalman filter implementation}}
\label{Alg:ExtendedKalmanFilter}
\begin{multicols}{2}
\begin{algorithmic}[1]
\System
\State $\bm{\dot{x}} = \bm{f} ( \bm{x}, t) + \bm{w};$
\State $\bm{y}_k = \bm{h} \, (\bm{x}_k) + \bm{v}_k;$ 
\vspace{0.2cm}
\Noise
\State $\bm{w} \sim (\bm{0}, \, \bm{Q});$
\State $\bm{v}_k \sim (\bm{0}, \, \bm{R}_k);$
\vspace{0.2cm}
\Initialization
\State $\bm{\hat x}_0^+ = E [ \bm{x}_0 ];$
\State $\bm{P}_0^+= E [ ( \bm{x}_0 - \bm{\hat x}_0^+)( \bm{x}_0 - \bm{\hat x}_0^+)^T];$
\Filter 
\vspace{0.2cm}
\item[\qquad \quad \textbf{Propagation}:]
\State $\bm{\dot {\hat{x}}} = \bm{f} (\bm{\hat x}, t);$
\State $\bm{\dot{P}} = \bm{F}\bm{P} + \bm{P} \bm{F}^T + \bm{Q};$ 
\vspace{0.2cm}
\item[\qquad \quad \textbf{Update}:]
\State $\bm{K}_k = \bm{P}_k^- \, \bm{H}^T ( \bm{H} \, \bm{P}_k^- \, \bm{H}^T + \bm{R}_k)^{-1};$
\State $\bm{\hat x}_k^+ = \bm{\hat x}_k^- + \bm{K}_k \, [\bm{y}_k - \bm{h} \, (\bm{\hat x}_k^-)];$
\State $\bm{P}_k^+ = (\bm{I} - \bm{K}_k \, \bm{H}) \, \bm{P}_k^-;$
\vspace{0.15cm}
\end{algorithmic}
\end{multicols}
\end{algorithm}

\section{Preliminary Results} \label{sec:Performances}
A three-object observation case is considered to assess the performances of the position determination method, while a single-object at a time tracking is devised for the Kalman filter application. 

\subsection{Three-object observation case}
The spacecraft is assumed to be on a heliocentric orbit in the ecliptic J2000 reference frame influenced by the gravity of the Sun only, from where it tracks three planets, namely Venus, Earth, and Mars. The initial epoch $t_0$ is set to 20$^{\textrm{th}}$ January 2020 00:00, and the initial state vectors of the bodies are shown in Table \ref{tab:SV0}, where the position components ($x_0$, $y_0$, $z_0$) are expressed in km and velocity components ($\dot{x}_0$, $\dot{y}_0$, $\dot{z}_0$) in km/s. \\[0.1cm]
\begin{table}[ht]
	\caption{Spacecraft and objects initial state vectors.}
	\label{tab:SV0}
	\centering
		\begin{tabular}{c c c c c c c}
			\specialrule{.10em}{.05em}{.05em}
			\specialrule{.10em}{.05em}{.05em}
& $x_0$ [km]& $y_0$ [km]& $z_0$ [km]& $\dot{x}_0$ [km/s]& $\dot{y}_0$ [km/s]& $\dot{z}_0$ [km/s]\\
\hline
S/C & -77484699,014 & 1.44753654,801 & -7097,387 & -32,392 & -15,471 & 0,0017 \\
Venus & 88620400,317 & 62344330,965 & -4303824,928 & -19,941 & 28,720 & 1,544 \\
Earth & -72168239,416 & 129721648,698 & -1881,250 & -26,540 & -14,596 & 0,002 \\ 
Mars & -171877932,528 & -159110369,541 & 8.49437,731 & 17,446 & -15,623 & -0,755 \\
			\specialrule{.10em}{.05em}{.05em}
			\specialrule{.10em}{.05em}{.05em}
		\end{tabular}
\end{table}

\begin{figure}[ht]
	\centering
	\includegraphics[width=0.8\columnwidth]{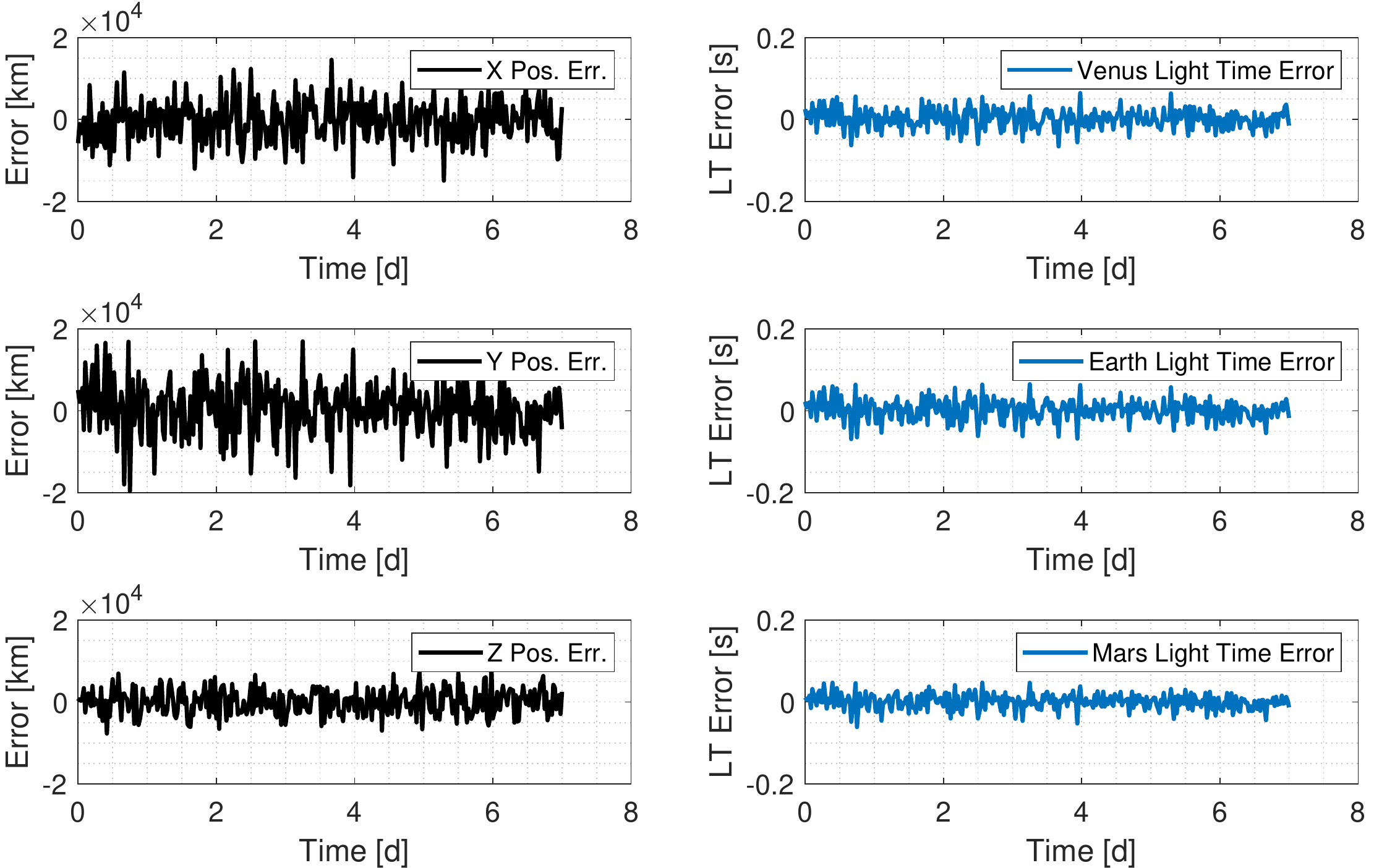}
	\caption{Position and light-time delay errors.}
	\label{fig:PosDet}
\end{figure}

The errors in position components and light-time delays to the tracked objects as output of the position determination method are shown in Figure \ref{fig:PosDet}. The apparent line-of-sight directions have been assumed to be affected by a 15 arcsecond noise (9 arcsecond for the attitude uncertainty deriving from orthogonal star trackers, 1 arcsecond for image processing, and 5 arcseconds accounted for thermo-mechanical errors and margins). In this case, the position components have been determined with an accuracy of 20,000 km and the light-time delays with an accuracy of 0.2 s. 
%
%

\subsection{Deep-Space CubeSat mission case}
\begin{figure}[ht]
     \centering
     \subfloat[Trajectory to Target 1.]{\includegraphics[width=0.3\textwidth]{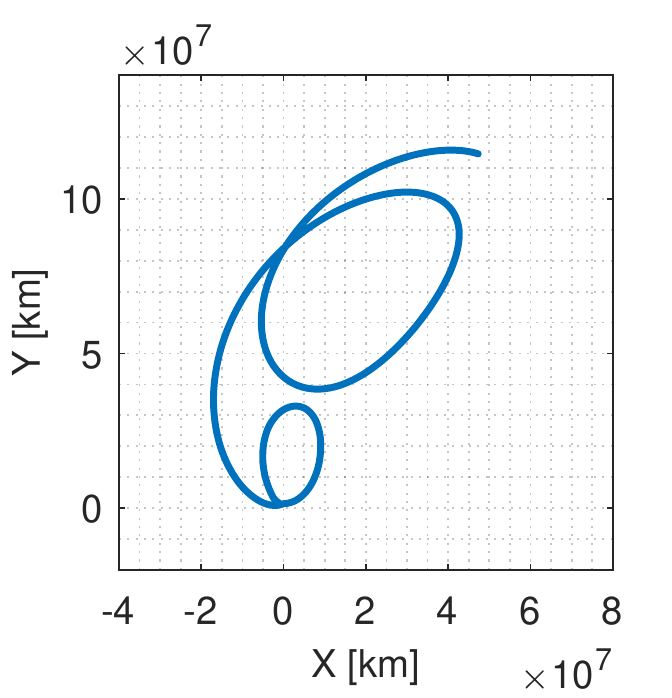}} \hspace{0.1cm}
     \subfloat[Trajectory to Target 2.]{\includegraphics[width=0.3\textwidth]{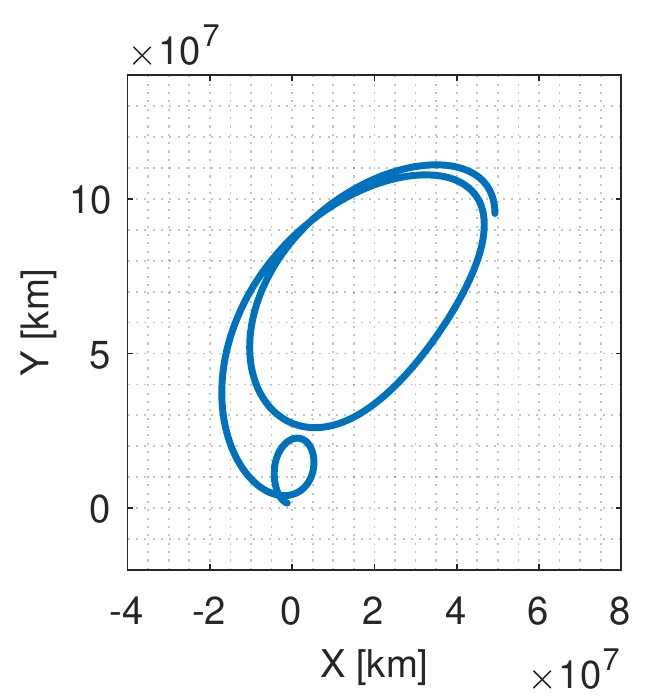}} \\ \vspace{-0.2cm}
     \subfloat[Trajectory to Target 3.]{\includegraphics[width=0.3\textwidth]{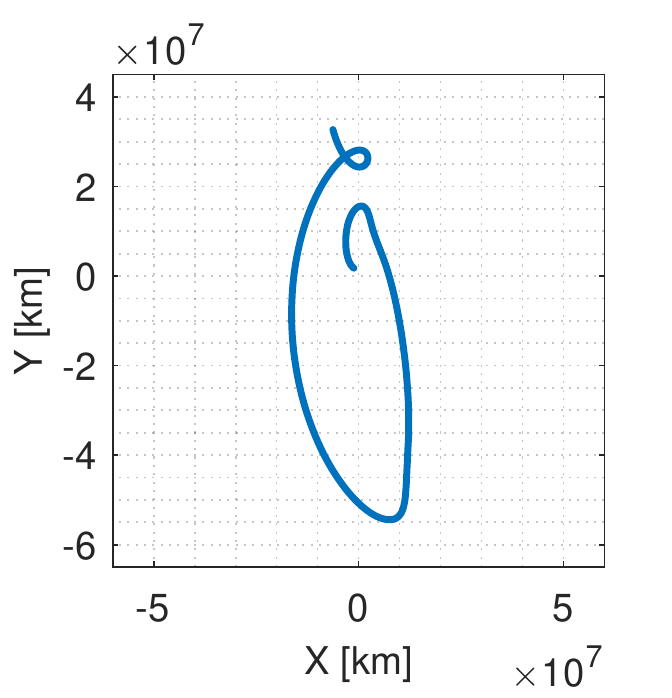}} \hspace{0.1cm}
     \subfloat[Trajectory to Target 4.]{\includegraphics[width=0.3\textwidth]{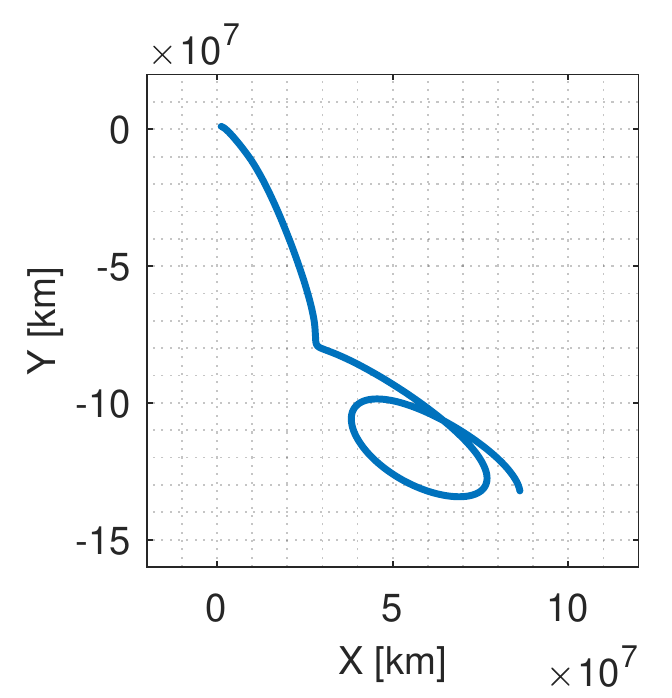}}
     \caption[Trajectories to four different targets.]{Trajectories to four different targets in the GSE frame.}
     \label{fig:MARGOTrajectories}
\end{figure}

In this Section, a Deep-Space CubeSat mission is considered. This sample mission aims to reach four different targets in the Solar System. The transfer trajectories are shown in Figure \ref{fig:MARGOTrajectories}. The extended Kalman filter shown in Section \ref{sec:Methodology} is implemented to the this mission case to assess the accuracy of the autonomous optical navigation exploiting line-of-sight observations. Owing to the characteristics of typical optical camera and star trackers for CubeSats, the probe will acquire the line-of-sight directions of one planet at a time for a given acquisition window, then slew to acquire another planet and feed the Kalman filter.
\subsubsection{Kalman filter settings} \label{subsec:kalmanfiltersettings}
The initial covariances of the position, velocity, and light-time delays have been set to
\begin{equation}
    \bm{P}_{r,0} = \sigma_r^2 \, \, \bm{I}_3 ;    \quad \qquad \bm{P}_{v,0} = \sigma_v^2 \, \, \bm{I}_3 ;    \quad  \qquad    \bm{P}_{\delta t,0} = \sigma_{\delta t}^2 \, \, \bm{I}_3 ;
\end{equation}
where $\sigma_r = 10^{5} \, \textrm{km}$, $\sigma_v = 10^{-1} \, \textrm{km/s}$, and $\sigma_{\delta t} = \sigma_r \textrm{/} c$. Thus, the initial state error covariance is the (6+i) dimensional matrix 
\begin{gather}
    \bm{P}_0 = \textrm{diag}(\bm{P}_{r,0} \,, \bm{P}_{v,0} \,, \bm{P}_{\delta t,0})
\end{gather}
where $\textrm{diag}(\cdot)$ is a diagonal matrix. The process noise covariance matrix has been set to the full matrix
\begin{equation}
    \bm{Q} = 10^{-12} \, \bm{1}_{6+i}           
\end{equation}
where $\bm{1}_{6+i} $ is a full matrix of ones with dimension ${6+i}$.
The apparent line-of-sight directions are affected by a white noise having a $3\sigma$ standard deviation of 15 arcseconds. The measurement covariance matrix is then 
\begin{equation}
    \bm{R} = \left(\frac{15/3}{3600} * \frac{\pi}{180} \right)^2 \, \, \bm{I}_2;           
\end{equation}
\subsubsection{Performances}
\begin{figure}[ht]
	\vspace{-0.6cm}
     \centering
	\includegraphics[width=0.8\textwidth]{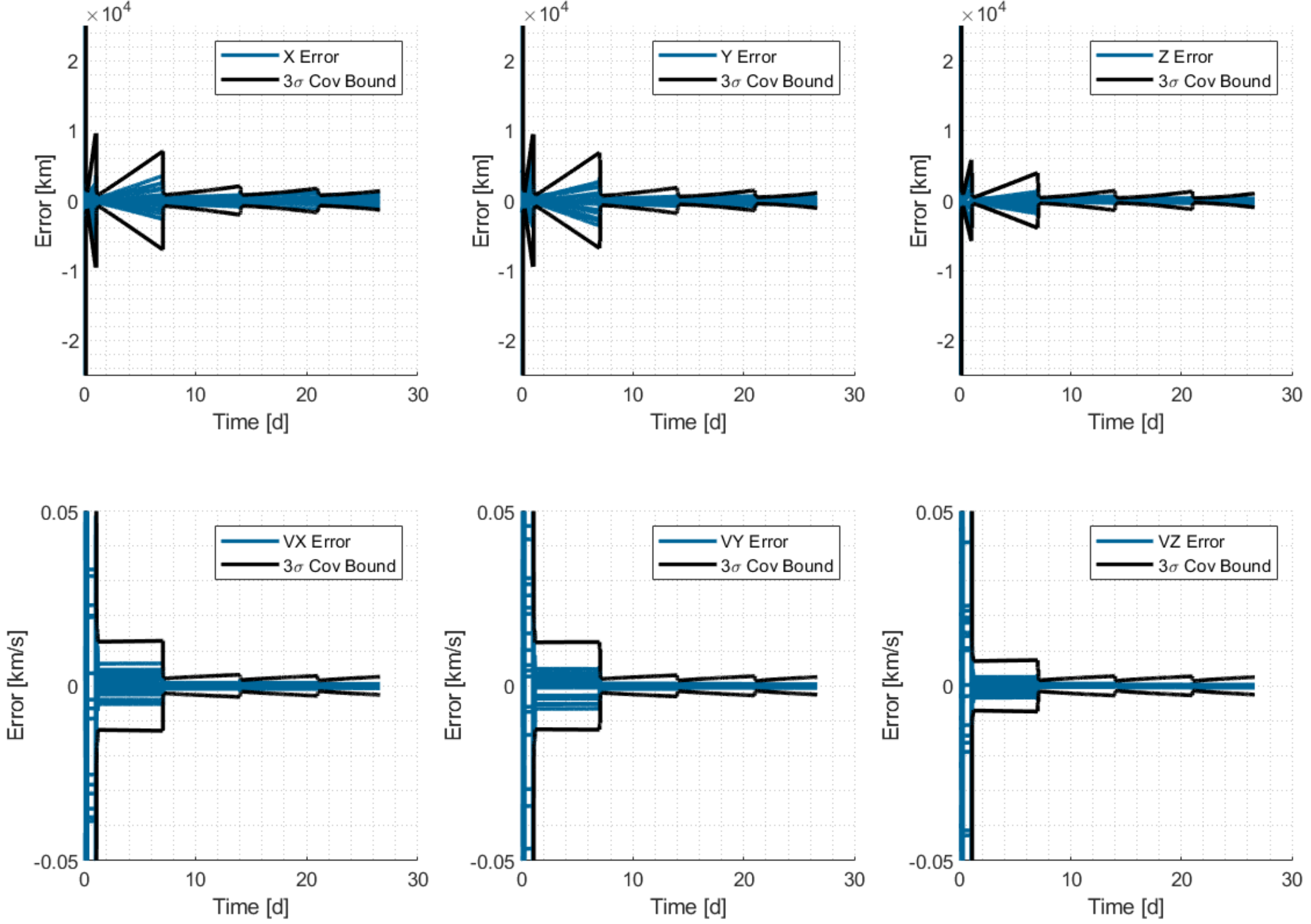}
    \caption[Kalman filter output at 0.1 Hz.]{Position and velocity error and covariance bounds as output of the Kalman filter with an acquisition frequency of 0.1 Hz for a portion of the spacecraft trajectory to Target 1, while tracking one planet per time (Earth and Mars) and accounting for the light-time delay to the planets. The slewing time (20 minutes) needed between the tracking periods is also considered.}  
    \label{fig:EKF_Slew}
\end{figure}

\begin{figure}[ht]
     \centering
	\includegraphics[width=0.5\textwidth]{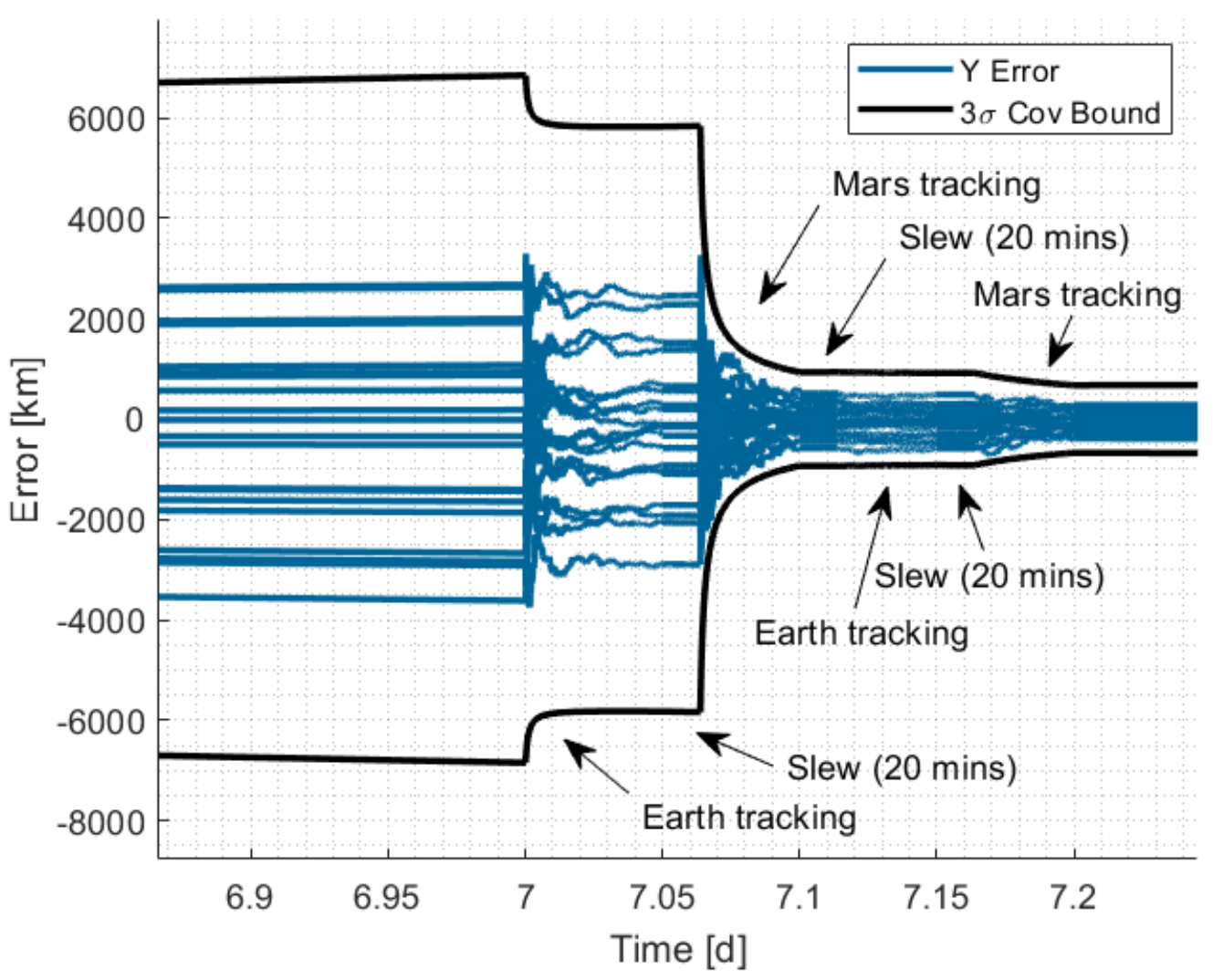}
    \caption[Kalman filter output at 0.1 Hz.]{Zoom on the planets tracking window. After the orbit propagation, the spacecraft points the Earth and acquires its LOS direction at a rate of 0.1 Hz for 1.2 h. Then, the probe slews towards Mars (20 minutes max) and tracks its LOS for another window of 1.2h at 0.1 Hz. Then the navigation cycle is repeated again. In overall, this navigation campaign lasts 3h.}  
    \label{fig:EKF_Slew_Zoom}
    \vspace{-0.6cm}
\end{figure}

The position and velocity errors and covariance bounds as output of the Kalman filter described in Section \ref{sec:Methodology} are shown in Figure \ref{fig:EKF_Slew}. In this scenario, the spacecraft orbit is propagated for 6 days and then 1 day is allocated for operations. In this day, the probe tracks one navigation beacon at a time, namely the Earth and Mars. After convergence, the position and velocity covariance bounds are lower than 1000 km and 2 m/s, respectively. In this simulation the slewing time required between the tracking windows is also accounted for and is estimated to be maximum 20 minutes for a 180 deg slew. The zoom of the acquisition window is shown in Figure \ref{fig:EKF_Slew_Zoom}. The spacecraft tracks the Earth from day 7 to day 7.05, then a slewing time is considered (till day 7.06) to point Mars, then the Mars tracking goes till day 7.1. Then another navigation cycle considering the slewing times between the targets is repeated. This navigation campaign lasts a total of 3h while tracking two different beacons one at a time.
\vspace{-0.2cm}
\section{Conclusions}
The feasibility of autonomous optical navigation for deep-space CubeSats has been investigated. The peculiar case of a deep-space cubesat is considered. To this purpose, an extended Kalman filter has been adopted to process the line-of-sight directions to the visible solar system bodies. The error in the line-of-sight directions is based on the performances of the cubesat sensors. The simulation scenario takes into account also the light-time delay to the navigation beacons and the slewing time required between the
targets tracking. The results in terms of position and velocity components estimations confirm the preliminary feasibility of autonomous optical navigation for deep-space CubeSat missions. 

\bibliography{main}

\end{document}